


\voffset 0.1in
\hoffset 0.1in

\input epsf

\def\square{\kern1pt\vbox{\hrule height 1.2pt%
\hbox{\vrule width 1.2pt\hskip 3pt\vbox{\vskip 6pt}%
\hskip 3pt\vrule width 0.6pt}\hrule height 0.6pt}\kern1pt}

\def\nolabels{\def\eqnlabel##1{}\def\eqlabel##1{}\def\reflabel##1{}}
\def\writelabels{\def\eqnlabel##1{%
{\escapechar=` \hfill\rlap{\hskip.09in\string##1}}}%
\def\eqlabel##1{{\escapechar=` \rlap{\hskip.09in\string##1}}}%
\def\reflabel##1{\noexpand\llap{\string\string\string##1\hskip.31in}}}
\nolabels

\global\newcount\secno \global\secno=0
\global\newcount\meqno \global\meqno=1

\def\eqn#1#2{\xdef #1{(\the\meqno)}\global\advance\meqno by1%
$$#2\eqno#1\eqlabel#1$$}

\def\refnn#1#2{\nref#1{#2}}
\def\nref#1#2{\xdef#1{\the\refno}%
\ifnum\refno=1\immediate\openout\rfile=refs.tmp\fi%
\immediate\write\rfile{\noexpand\item{#1\ }\reflabel{#1}#2}%
\global\advance\refno by1}
\def\addref#1{\immediate\write\rfile{\noexpand\item{}#1}}

\global\newcount\refno \global\refno=1
\newwrite\rfile
\def\ref#1#2{\the\refno\nref#1{#2}}
\def\nref#1#2{\xdef#1{\the\refno}%
\ifnum\refno=1\immediate\openout\rfile=refs.tmp\fi%
\immediate\write\rfile{\noexpand\item{#1.\ }\reflabel{#1}#2}%
\global\advance\refno by1}
\def\addref#1{\immediate\write\rfile{\noexpand\item{}#1.}}

\headline={\ifnum\pageno=1\firstheadline\else
\ifodd\pageno\rightheadline \else\leftheadline\fi\fi}
\def\firstheadline{\hfil}
\def\rightheadline{\hfil}
\def\leftheadline{\hfil}

\footline={\ifnum\pageno=1\firstfootline\else\otherfootline\fi}
\def\firstfootline{
\vbox{\baselineskip=11pt
  \vskip 7pt
  \hbox to \hsize{\hrulefill}
  \hbox to \hsize{{\sl Third Workshop on Thermal Field Theories}
                  \hss {\tenrm Banff, Canada ~(August 1993)}
                  \hss {\tenrm MIT-CTP-2254}}
  \vskip 1pt
  \hbox to \hsize{\ninerm hep-ph/9311324. ~~ This work is supported in part
                  by funds provided by the U.S.~Department of Energy}
  \hbox to \hsize{\ninerm (D.O.E.) under contract \#DE-AC02-76ER03069, and
         in part by {\ninesl Istituto Nazionale di Fisica Nucleare}, Italy.}}}

\def\otherfootline{\hfil}

\font\twelvebf=cmbx10 scaled\magstep 1
\font\twelverm=cmr10 scaled\magstep 1
\font\twelveit=cmti10 scaled\magstep 1

\font\elevenit=cmti10 scaled\magstephalf

\font\tenbf=cmbx10
\font\tenrm=cmr10
\font\tenit=cmti10

\font\ninerm=cmr9

\font\ninesl=cmsl9

\parindent=1.5pc
\hsize=6.0truein
\vsize=8.5truein

\centerline{\tenbf FIELD THEORETICAL BACKGROUND FOR THERMAL PHYSICS}
\vglue 0.8cm
\centerline{\tenrm Roman JACKIW}
\baselineskip=13pt
\centerline{\tenit Center for Theoretical Physics, Laboratory
for Nuclear Science,}
\baselineskip=12pt
\centerline{\tenit and Department of Physics,
Massachusetts Institute of Technology}
\baselineskip=12pt
\centerline{\tenit Cambridge, Massachusetts 02139, USA}
\vglue 0.3cm
\centerline{\tenrm notes by}
\vglue 0.3cm
\centerline{\tenrm Giovanni AMELINO-CAMELIA}
\baselineskip=13pt
\centerline{\tenit Center for Theoretical Physics, Laboratory
for Nuclear Science,}
\baselineskip=12pt
\centerline{\tenit and Department of Physics,
Massachusetts Institute of Technology}
\baselineskip=12pt
\centerline{\tenit Cambridge, Massachusetts 02139, USA}
\vglue 0.8cm
\centerline{\tenrm ABSTRACT}
\vglue 0.3cm
{\rightskip=3pc
 \leftskip=3pc
 \tenrm\baselineskip=12pt\noindent
Techniques of zero-temperature field theory
that have found application in the analysis of field theory at finite
temperature are revisited.
Specifically, several of the results that are discussed
are relevant
to the study of symmetry-changing phase transitions and high temperature
QCD, which today are among the most actively investigated problems
in finite temperature field theory.
\vglue 0.6cm}

\vfil
\twelverm
\baselineskip=14pt
\leftline{\twelvebf 1. Overview}
\vglue 0.4cm
Most of the techniques used in finite temperature field theory were first
developed for analysis of field theory at zero temperature.
The purpose of these pedagogical lectures is to
revisit (a part of) this field theoretical background.

We start, in Secs.~2 and 3,
by discussing techniques for studying the phase
structure of a field theory using Green's
functions generating functionals.
These techniques are receiving renewed interest
because of their role in numerous recent studies
of temperature-induced symmetry-changing
phase transitions, which have followed the suggestion that the
observed baryon number asymmetry might have originated at the electroweak
phase transition.

Also relevant to the investigation of
temperature-induced phase transitions is the formalism
of non-equilibrium quantum field theories,
which we discuss in Sec.4.

Finally, in Sec.~5 we review Chern-Simons theory, which has recently
found application in
finite temperature field theory because the Chern-Simons
eikonal gives the generating functional of
the {\elevenit hard thermal loops} in QCD.

\vglue 0.6cm
\leftline{\twelvebf 2. Phase Structure and
Generating Functionals }
\vglue 0.3cm
\leftline{\twelveit 2.1. Effective Action}
\vglue 1pt
An important aspect of a physical
system is its phase structure (what are the available phases,
in what phase the system is at a given time,
what is the nature of the phase transitions ....).
Let us assume, at least at the beginning, that one of the fields
in the theory ($\Phi$) is a scalar and study
the phase structure in terms
of the thermal expectation of that scalar field,
defined by (we use units such that $\hbar=c=k_{Boltzman}=1$)
\eqn\ev{\eqalign{<\Phi> \equiv {tr ~ e^{- \beta H} \Phi \over
tr ~ e^{- \beta H}} ~,}}
where $\beta$ is the inverse temperature $1/T$.
At zero temperature one is
interested in the zero-temperature limit of $<\Phi>$,
{\elevenit i.e.} the vacuum expectation value $<0| \Phi |0>$.
If one could solve the theory completely these expectation values could
be evaluated explicitly from their definitions.
Unfortunately there are
only a few quantum field theories that can be solved completely
(and they are
not very interesting from the physical point of view ),
so that the answer to the problem at hand can only
be approached somewhat indirectly by setting up additional formalism,
with the hope that a resonable approximation scheme emerges.

We use an approach based on
generating functionals:
the effective action, the effective energy, and the effective potential.
This was first proposed by Jona-Lasinio$^{\ref\jona{G. Jona-Lasinio,
Nuovo Cimento {\bf 34} (1964) 1790.}}$,
and later developed in Refs.[$\ref\colewei{S. Coleman and E. Weinberg,
Phys. Rev. {\bf D7} (1973) 1888.}\!-
\refnn\jakdi{R. Jackiw, Phys. Rev. {\bf D9} (1974) 1686.}
\refnn\doja{L. Dolan and R. Jackiw,
Phys. Rev. {\twelvebf D 9}, 3320 (1974).}
\!\ref\wei{S. Weinberg, Phys. Rev. {\twelvebf D 9}, 3357 (1974).}$].
(For a review of this subject
see Ref.[$\ref\rive{R. J. Rivers, {\twelveit Path Integrals Methods
in Quantum
Field Theory} (Cambridge University Press, Cambridge, 1987).}$].)
For definiteness,
we work at zero-temperature; however, the finite temperature
generalizations can be easily obtained following the discussion in Sec.3.

The first step is the introduction of the ``partition function" $Z(J)$
\eqn\zusu{\eqalign{Z(J)   \equiv
 <0| {\cal T} \exp i \int_x J(x) \Phi(x) |0> ~,}}
where ${\cal T}$ is the time-ordering operator.

$Z(J)$ is a derived quantity, a quantity external to the theory,
but its importance is due to the fact that $Z(J)$
is the generating functional for Green's functions:
\eqn\aaaa{\eqalign{G_n(x_1,.....,x_n) \equiv
<0| {\cal T} \Phi(x_1) ..... \Phi(x_n) |0>   =  \biggl[{\delta^n
\over i^n \delta J(x_1) .... J(x_n)} Z(J) \biggr]_{J=0} ~.}}
In as much as all the physical information on a quantum field theory
is encoded in
the Green's functions, one can study the theory just using
the generating functional $Z(J)$.

The generating functional for connected Green's functions $W(J)$
is given by
\eqn\wusu{\eqalign{W(J)  \equiv - i \ln Z(J) ~ .}}
The same physical information encoded in the Green's functions
is organized more economically
in the connected Green's functions; in fact,
once the connected ones are known the complete Green's functions
can be reconstructed.

The effective action $\Gamma(\phi)$ is
defined as the functional Legendre transform
of $W(J)$:
\eqn\gusu{\eqalign{\Gamma(\phi) = W(J) -
\int_x ~ \phi(x) J(x)  ~, }}
where
\eqn\fidef{\phi(x) \equiv {\delta W(J) \over \delta J(x)} ~, }
and in \gusu\space $J$ is expressed in terms of $\phi$ using \fidef.
[$\Gamma(\phi)$ is sometimes referred to as the ``quantum action";
in the limit $\hbar \rightarrow 0$ it reproduces the
classical action.]

Evidently
\eqn\statjoja{{\delta \Gamma(\phi) \over \delta \phi(x)} = - J(x) ~, }
and, as a consequence, physical solutions, which require
vanishing $J$, correspond to the stationary points of $\Gamma$.

 From the definitions that we have introduced
it is clear that
\eqn\vev{<0| \Phi(x) |0> = [\phi(x)]_{J=0} \equiv \phi_0(x) ,}
and therefore using Eq.\statjoja\space
one finds that the vacuum expectation value of $\Phi$
is given by the value of $\phi(x)$ which stationarizes $\Gamma(\phi)$:
\eqn\vevfio{\biggl[{\delta \Gamma(\phi)
\over \delta \phi(x)}\biggr]_{\phi(x)=\phi_0(x)} = 0
\rightarrow \phi_0(x) = <0| \Phi(x) |0>
{}~. }
We have replaced the problem of evaluating the
vacuum expectation value of a quantum field by the
problem of finding the stationary points of a classical functional.
Such a problem has a long tradition in physics and we can use
the numerous approximation
techniques that have been developed over the years.

Finally we note that $\Gamma(\phi)$ is  the
generating functional of the one particle
irreducible\footnote{*}{\ninerm\baselineskip=11pt
A graph is said to be ``one particle irreducible" if it does not
become disconnected upon opening any one line; otherwise it is
``one particle reducible".}
(1PI) Green's functions:
\eqn\opi{\biggl[{\delta^n \Gamma(\phi)
\over \delta \phi(x_1) ..... \delta \phi(x_n)}\biggr]_{\phi(x)=\phi_0(x)}
= G_n^{1PI}(x_1,.....,x_n) ~. }
These contain all the physical information
encoded in the Green's functions in an
even more economical format than the connected Green's functions.

\vglue 0.3cm
\leftline{\twelveit 2.2. Effective Potential and Effective Energy}
\vglue 1pt
If one is interested in all the physical
information in a quantum field theory,
it is necessary to study the complete effective action, as a functional
of the arbitrary background field $\phi(x)$.  However, when one is
only interested in the vacuum expectation value $\phi_0(x)=<0| \Phi(x) |0>$,
one can use any {\elevenit a priori}
information available
on $\phi_0(x)$ to simplify the analysis.

If it is expected
that translation invariance is not broken, as is usually the case,
one can assume that $\phi_0(x)$ be
independent of the space-time point: $\phi_0(x)=constant=\phi_0$.
With this hypothesis, in order to determine $\phi_0$ it is sufficient to
study the effective action as a function of a constant background $\phi$.
However, when $\phi$ is constant an infinite
volume factor arises from the space-time integrations, and
therefore,
rather than working with the effective action,
we introduce the effective potential $V(\phi)$,
defined by
\eqn\vdef{V(\phi) \int_x \equiv
- [\Gamma(\phi)]_{\phi=constant} ~.}

The effective potential is the generating functional of
the 1PI Green's functions at zero energy and
momentum\footnote{*}{\ninerm\baselineskip=11pt Note
that in general the Green's functions at zero energy and momentum are
not physical,
though they might have physical relevance in particular contexts.
The poles and the residues at the poles of the
Green's functions are physical.},
and in the $\hbar \rightarrow 0$ limit the effective potential reproduces
the classical potential.
Clearly from Eqs.\vevfio\space and \vdef\space it follows that for
translationally invariant theories the vacuum expectation value
can be determined using the effective potential:
\eqn\veffio{\biggl[{\delta V(\phi)
\over \delta \phi}\biggr]_{\phi=\phi_0} = 0
\rightarrow \phi_0 = <0| \Phi(x) |0>
{}~. }

An intermediate possibility between evaluating the
effective action as a functional of a space-time dependent
background $\phi(x)$ and the much simpler case of a constant background,
is the evaluation of the effective action for static (time-independent,
position-dependent) background $\phi({\bf x})$.
In this circumstance, the time
integration in $\Gamma(\phi)$
leads to an overall (infinite)
time-interval factor;
it is therefore convenient to
introduce the effective energy $E(\phi)$,
defined by
\eqn\gggg{E(\phi) \int dx_0 \equiv - [\Gamma(\phi)]_{\phi=static} ~.}

The effective energy is the generating functional
for 1PI Green's functions at zero energy,
in the $\hbar \rightarrow 0$ limit it reproduces
the classical energy, and in time translation-invariant theories
it can be used to determine the vacuum expectation value using:
\eqn\eeffio{\biggl[{\delta E(\phi)
\over \delta \phi({\bf x})}\biggr]_{\phi({\bf x})=\phi_0({\bf x})} = 0
\rightarrow \phi_0({\bf x}) = <0| \Phi(x) |0>
{}~. }

\vglue 0.3cm
\leftline{\twelveit 2.3. Loop Expansion}
\vglue 1pt
In the search for the stationary points of $\Gamma(\phi)$, the choice
of a strategy
for evaluating $\Gamma(\phi)$ is very important.
One commonly used technique is the approximation of $\Gamma(\phi)$
based on the following known$^{\jakdi,\doja}$
expansion\footnote{**}{\ninerm\baselineskip=11pt
The expansion is formally in powers of
$\hbar$
but this doesn't explicitly appear in Eq.(15) because in our units
$\hbar=1$.}
\eqn\gexpusu{\Gamma(\phi)=I(\phi)-{i \over 2} tr \ln D
+ \Gamma_2(\phi)  ~.}
$I(\phi)$ is the classical action.
The functional operator $D^{-1}(\phi;x,y)$ is defined by:
\eqn\dfunct{D^{-1}(\phi;x,y) =
 {\delta^2 I(\phi) \over \delta \phi(x) \delta \phi(y)} ~.}

\eject

\noindent
The term ${i \over 2} [ tr \ln D]$,
where the trace and the logarithm are taken in the functional
sense,
is the one loop contribution [it is $O(\hbar)$].
$\Gamma_2(\phi)$ is computed as follows.
In the
classical action $I(\Phi)$, shift the quantum field $\Phi(x)$ by a
c-number field $\phi(x)$. $I(\Phi+\phi)$ contains terms cubic and higher
in $\Phi$ that define an interaction part $I_{int}(\phi;\Phi)$
whose vertices depend on $\phi$.
$\Gamma_2(\phi)$ is the sum of all the two-or-more-loop 1PI vacuum
graphs in the theory with vertices given by $I_{int}(\phi;\Phi)$ and
propagators taken to be $D(\phi;x,y)$.

For example, for the $\lambda \Phi^4$ scalar theory
\eqn\if{I(\Phi) =
\int_x \biggl[
{1 \over 2} (\partial_{\mu} \Phi) (\partial^{\mu} \Phi)
+{a \over 2}  \Phi^2
-{\lambda \over 4!}  \Phi^4 \biggr]
{}~,}
\eqn\dfif{D^{-1}(\phi;x,y) =
(- \square + a - {\lambda \over 2} \phi^2(x) ) \delta^4(x-y) ~,}
\eqn\iin{I_{int}(\phi;\Phi) = \int_x \biggl[
- {\lambda \over 6} \phi \Phi^3 - {\lambda \over 4!} \Phi^4 \biggr]~.}
The diagrams which contribute to $\Gamma_2(\phi)$ for the $\lambda \Phi^4$
scalar theory are shown in Fig.1 up to three loops.
The lines represent $D(\phi;x,y)$ and, as indicated
by Eq.\iin, there are three-point and four-point vertices.

\midinsert
\epsfxsize=4in
\centerline{\epsffile{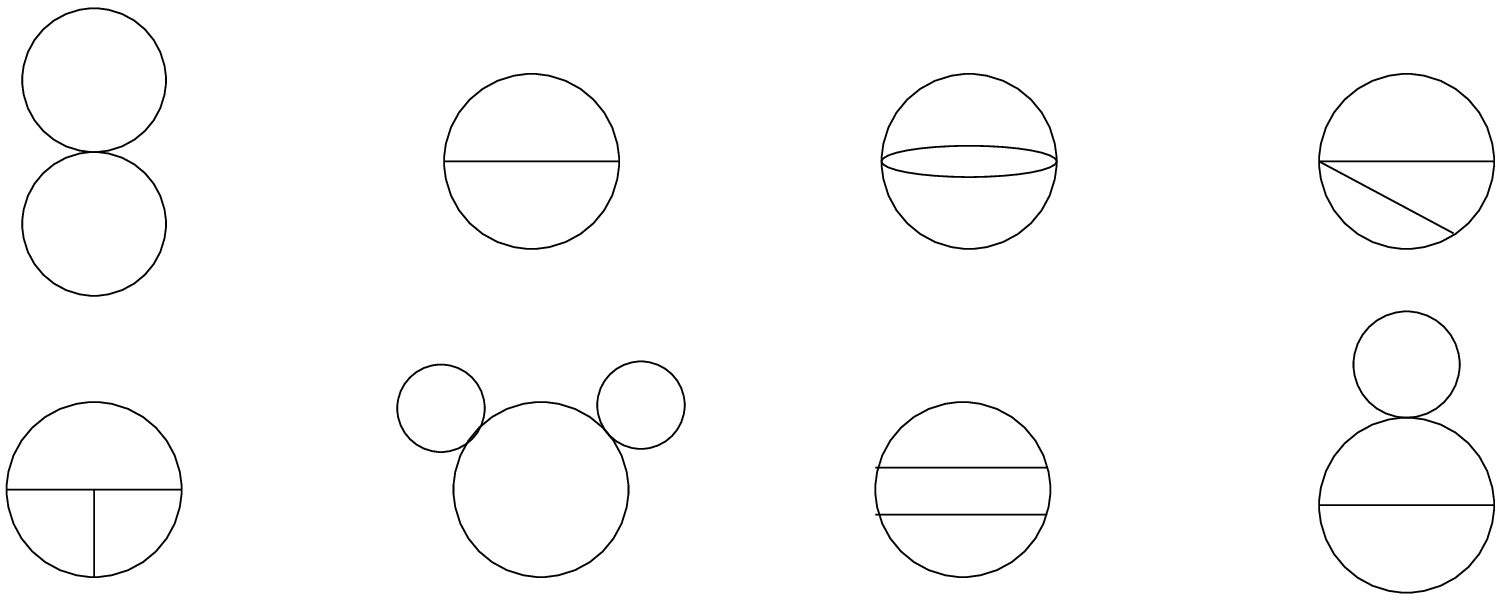}}
\bigskip
\centerline{Fig.~1}
\endinsert

A loop expansion for the
effective potential can obviously be obtained by evaluating Eq.\gexpusu\space
for constant $\phi$ and extracting a (infinite) volume factor:
\eqn\vexpusu{V(\phi)=V_{cl}(\phi)+{i \over 2} \int_k
\ln D(\phi;k) + V_2(\phi)  ~,}
where $V_{cl}(\phi)$ is the classical potential,
$D(\phi;k)$ is the Fourier-transform of $D(\phi;x,y)$ at constant $\phi$
\eqn\dftr{D(\phi;x,y) \equiv
\int_k e^{-i {\bf k} \cdot ({\bf x}-{\bf y})}  D(\phi;{\bf k}) ~,}
and $V_2(\phi)$ is given by
\eqn\vtwodef{V_2(\phi) \int_x \equiv - [\Gamma_2(\phi)]_{\phi=constant} ~.}

Analogously, the loop expansion for the effective energy is obtained by
evaluating Eq.\gexpusu\space for static $\phi$ and extracting a (infinite)
time-interval factor.

\eject

\noindent

\vglue 0.3cm
\leftline{\twelveit 2.4. Selective Summations and
Effective Action for Composites}
\vglue 1pt
It is very difficult to carry out
the loop expansion beyond two loops.
It is
however possible to
perform selective summations of higher loop graphs.
One way to systematize such summations is the large $N$ method, in which
one considers an $N$-component field and uses the fact that in
the large $N$ limit some multi-loop graphs give
greater contributions than others.
For example in the $\lambda \Phi^4$ $N$-component scalar theory the leading
multi-loop contributions come from {\elevenit bubble-graphs}
(see Fig.2).

\midinsert
\epsfxsize=2.5in
\centerline{\epsffile{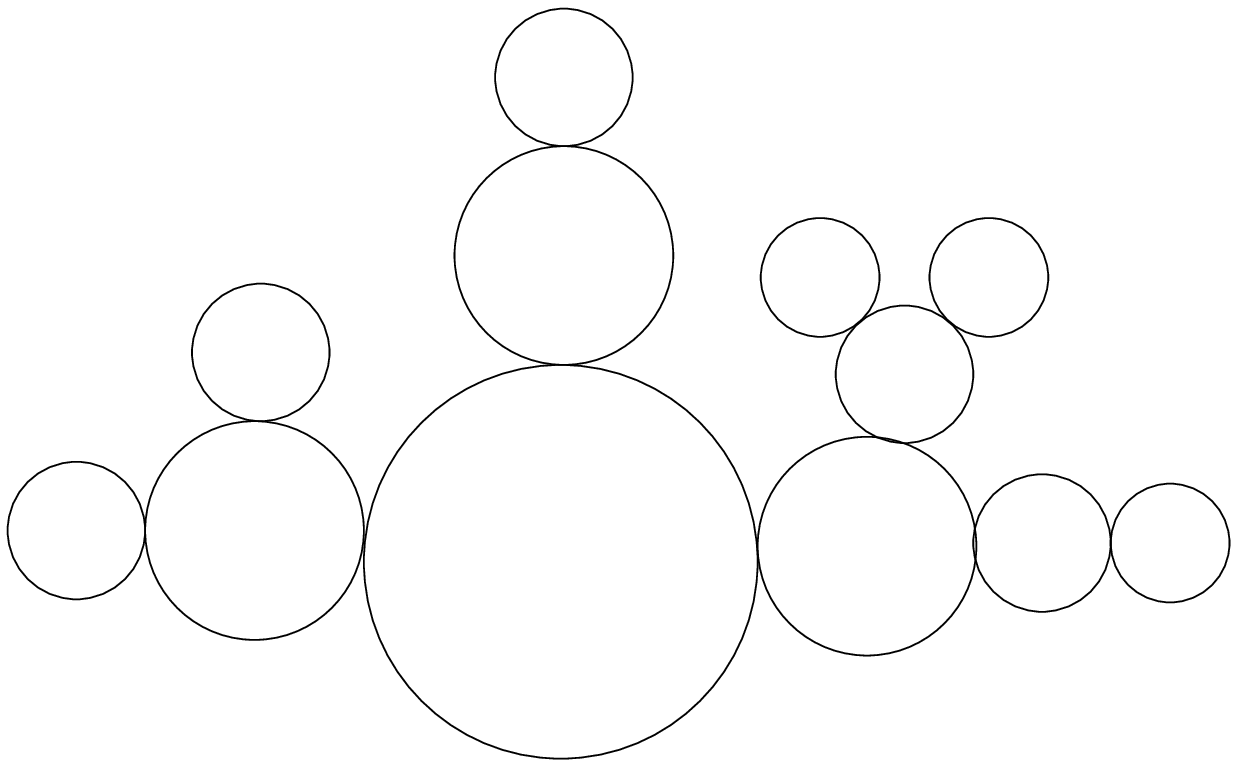}}
\bigskip
\centerline{Fig.~2}
\endinsert

Another way to perform systematic selective summations makes use of the
effective action for composites$^{\ref\corn{J.M. Cornwall,
R. Jackiw, and E. Tomboulis,
Phys. Rev. {\twelvebf D 10} (1974) 2428.}}$.
This generalization of the effective action was primarily
introduced to seek the phase structure at the level of the two-point
function\footnote{*}{\ninerm\baselineskip=11pt
For example Fermi fields do not have a vacuum expectation value, but  the
fermion propagator can show spontaneous chiral symmetry breaking through mass
generation.},
in addition to, or
instead of, the one-point function.
Moreover, it can also be used to
rexpress the ordinary effective action in a way that is more suitable
for selective summations.

One arrives at the definition of the effective action for composites
following a path very analogous to the one that leads to the
ordinary effective action.
The starting point is a new partition function that depends on a
bilocal source $K(x,y)$, in addition to the local source $J(x)$:
\eqn\zcjt{\eqalign{Z(J,K) \equiv e^{i W(J,K)}
\equiv <0| {\cal T} \exp i \biggl[ \int_x J(x) \Phi(x)+
{1 \over 2} \int_{x,y} \Phi(x) K(x,y) \Phi(y) \biggr] |0> ~,}}
where we have also introduced $W(J,K)$.
A possible expectation value of the quantum field $\Phi(x)$ is $\phi(x)$;
this is
again defined as the functional derivative of $W$ with respect to $J$
\eqn\fdef{ \phi(x) \equiv {\delta W(J,K) \over \delta J(x)} ~, }
but now one also introduces $G(x,y)$,
defined by
\eqn\gdef{{\delta W(J,K) \over \delta K(x,y)} \equiv
{1 \over 2} \phi(x) \phi(y) + {1 \over 2} G(x,y) ~, }
which is a possible expectation value
of the connected part of the two point function.

The effective action for composites
$\Gamma(\phi,G)$  is
defined as the double Legendre
transform of $W(J,K)$
\eqn\gamcjt{\eqalign{\Gamma(\phi,G) & = W(J,K) -
\int_x ~ \phi(x) J(x)  \cr
&~~ - {1 \over 2} \int_{x,y} \phi(x) K(x,y) \phi(y)
- {1 \over 2} \int_{x,y}  G(x,y) K(y,x) ~, }}
where $J$ and $K$ are to be expressed in terms of $\phi$ and
$G$ using \fdef\space and \gdef.

One can show$^{\corn}$
that $\Gamma(\phi,G)$ is the generating functional
in $\phi$ for two-particle
irreducible\footnote{*}{\ninerm\baselineskip=11pt
A graph is said to be ``two particle irreducible" if it does not
become disconnected upon opening any two lines; otherwise it is
``two particle reducible".} (2PI)
Green's functions, expressed in
terms of the full propagator $G$.

 From Eq.\gamcjt\space it is evident that
\eqn\invlega{{\delta \Gamma(\phi,G) \over \delta \phi(x)} =
- J(x) - \int_y ~K(x,y) \phi(y)  ~, }
\eqn\invlegb{{\delta \Gamma(\phi,G) \over \delta G(x,y)} =
- {1 \over 2} K(x,y) ~. }
Since
physical processes correspond to vanishing sources $J$ and
$K$, Eqns.\invlega\space and \invlegb\space imply that
$\Gamma(\phi,G)$ is physical at its stationary points.

Eq.\invlegb\space also indicates the
relation between $\Gamma(\phi,G)$ and the ordinary effective action;
in fact,
as it is clear from its definition, $\Gamma(\phi,G)$
must reproduce $\Gamma(\phi)$ when $K=0$, and
using Eq.\invlegb\space one finds that
\eqn\gusucjt{ \Gamma(\phi) = \Gamma(\phi,G_0) ~, }
when $G_0$ satisfies the stationary requirement
\eqn\gapori{\biggl[ {\delta \Gamma(\phi,G)
\over \delta G(x,y)}\biggr]_{G=G_0} = 0 ~. }

A loop expansion is also
available$^{\corn}$ for the
effective action for composites
\eqn\gexp{\Gamma(\phi,G)=I_{cl}(\phi)-
{i \over 2} tr [\ln G - D^{-1} G ]
+ \Gamma_2(\phi,G)  ~,}
where $\Gamma_2(\phi,G)$  is given by the sum of the two-or-more-loop
2PI vacuum
graphs in the theory with vertices given by $I_{int}(\phi;\Phi)$ and
propagators set equal to $G(x,y)$.

For $G=G_0$ Eq.\gexp\space gives a new expansion for the ordinary
effective action. Moreover, using Eq.\gexp\space the gap equation
\gapori\space which determines $G_0$ can also
be expressed in terms of a loop expansion as follows
\eqn\gap{G_0^{-1}(x,y) =  D^{-1}(x,y)
- 2 i \biggl[ {\delta \Gamma_2(\phi,G)
\over \delta G(x,y)} \biggr]_{G=G_0} ~.}
Eq.\gap\space gives the relation between $G_0$ and the {\elevenit tree-level
propagator} $D$, and can therefore be used to express the diagrams of
the expansion \gexp\space in terms of diagrams of the ordinary
expansion \gexpusu.
For example,
for the $\lambda \Phi^4$ scalar theory it is easy to show that
the summation of all the contributions of the
bubble graphs
to the ordinary effective action $\Gamma(\phi)$
corresponds to the solution of the Eqs.\gusucjt-\gapori\space when
$\Gamma_2(\phi,G)$ is approximated by its only graph of $O(\lambda)$,
{\elevenit i.e.}
the double-bubble graph in Fig.1.
This bubble-resummation has been used in a recent study of the
temperature-induced phase transition of the $\lambda \Phi^4$ scalar
theory$^{\ref\gacpi{G. Amelino-Camelia and
S.-Y. Pi, Phys. Rev. {\twelvebf D??} (1993) ????.}}$.
In general
the solution of Eqs.\gusucjt-\gapori\space with the
approximation of $\Gamma_2(\phi,G)$ by a finite sum
of its graphs corresponds to the selective summation of an infinite class of
``ordinary graphs" [graphs of the ordinary expansion \gexpusu].

When interested in translation-invariant (time translation-invariant)
solutions one can perform
selective summations of multi-loop graphs which contribute to the
effective potential (energy).
These summations may be performed systematically using the effective
potential for composites $V(\phi,G)$
and the effective energy for composites $E(\phi,G)$,
which, in analogy with Eqs.\vdef-\gggg,
are defined in terms of the effective action for composites by
\eqn\vcjtdef{V(\phi,G) \int_x \equiv
- [\Gamma(\phi,G)]_{translation-invariant} ~,}
\eqn\ecjtdef{E(\phi,G) \int dx_0 \equiv
- [\Gamma(\phi,G)]_{time \, translation-invariant} ~.}

\vglue 0.3cm
\leftline{\twelveit 2.5. Variational Interpretation and
Variational Approximations}
\vglue 1pt
We have seen that, in the formalism of these generating
functionals, physical solutions satisfy
variational equations which
stationarize the functionals.
In this subsection we shall discuss the fact
that the functionals themselves have a variational interpretation.
The knowledge of these variational interpretations is
important, especially because they can suggest approximations.

Let us start with the effective energy $E(\phi)$.
It can be shown that $E(\phi)$
is the minimal expectation value of the
Hamiltonian in normalized states which are constrained
so that the expectation value of the quantum field $\Phi(x)$ in those states
is equal to $\phi({\bf x})$, {\elevenit i.e.}
\eqn\evarint{\eqalign{E(\phi) =& min <\Psi| H |\Psi> ~,
\cr
<\Psi|\Psi> = 1 & ~, ~~ <\Psi| \Phi(x) |\Psi>  = \phi({\bf x})
{}~. }}
This result can be immediately related to the general quantum mechanical
principle that in searching for the solutions of
a static problem one should minimize the
expectation value of the Hamiltonian.
Using the effective energy this principle is implemented in two steps:
one first minimizes the expectation value of the Hamiltonian with the
constraint $<\Psi| \Phi |\Psi>  = \phi$,
and then the resulting
$\phi$-dependent functional (the effective energy)
is minimized with respect to $\phi$.

For the effective potential $V(\phi)$ a similar variational interpretation
and similar considerations apply; one finds that
\eqn\vvarint{\eqalign{V(\phi) =& min \biggl[ {<\Psi| H |\Psi>
\over \int d{\bf x}} \biggr]  ~,
\cr
<\Psi|\Psi> = 1 & ~, ~~ <\Psi| \Phi(x) |\Psi>  = \phi
{}~. }}

Also the effective potential and the effective energy for composites
have a variational interpretation.
For example to obtain $E(\phi,G)$ one minimizes the expectation value
of the Hamiltonian on normalized states which satisfy two constraints:
\eqn\ecjtvarint{\eqalign{E(\phi,G) =& min <\Psi| H |\Psi> ~,
\cr
<\Psi|\Psi> = 1~, ~&~ <\Psi| \Phi(x) |\Psi>  =  \phi({\bf x}) ~,
\cr
<\Psi| \Phi(x) \Phi(y) |\Psi>&|_{x_0=y_0} = \phi({\bf x}) \phi({\bf y})+
G({\bf x},{\bf y})
{}~. }}

The variational definitions for the effective potential and the
effective energy
are related to the general quantum mechanical variational
principle for static problems.
Correspondingly,
the variational definition for the effective action $\Gamma(\phi)$
can be derived from the
Dirac variational principle$^{\ref\timedep{A. Kerman and
R. Jackiw, Phys. Lett. {\twelvebf A71} (1979) 159;
R. Jackiw, Int. J. Quantum Chem. {\bf 17} (1980) 41.}}$
for time-dependent problems, which instructs
minimizing the following quantity
\eqn\dirac{\int dt <\Psi,t| i \partial_t - H |\Psi,t> \equiv
\int dt d{\bf x} \biggl[ \Psi^{*}(t,{\bf x})
(i \partial_t - H) \Psi(t,{\bf x}) \biggr] }
with the normalization constraint
\eqn\dirnor{<\Psi,t|\Psi,t> \equiv
\int d{\bf x} \Psi^{*}(t,{\bf x}) \Psi(t,{\bf x}) = 1 ~. }
$\Gamma(\phi)$ can be variationally defined as follows
\eqn\gdirac{\Gamma(\phi) =~ min \biggl[
\int dt <\Psi_-,t| i \partial_t - H |\Psi_+,t> \biggr] }
where the states $|\Psi_+,t>$ and $|\Psi_-,t>$ are constrained so that
\eqn\constr{\eqalign{<\Psi_-,t| \Phi({\bf x}) |\Psi_+,t>
&= \phi(t,{\bf x})
\cr
<\Psi_-,t|\Psi_+,t> &= 1
\cr
\lim_{t \rightarrow \pm \infty} |\Psi_\pm,t> &= |0>
{}~.}}

The variational definition of the
effective action for composites $\Gamma(\phi,G)$ is analogous, but with
the addition of a constraint
for $<\Psi_-,t| \Phi({\bf x}) \Phi({\bf y}) |\Psi_+,t>$.

The variational interpretation of our generating functionals naturally
leads to some variational approximations.
These approximations are best formulated
in the Schr\"{o}dinger picture for quantum
field theory$^{\ref\schrpict{R. Jackiw,
Canad. Math. Soc. Conference Proceedings
{\twelvebf 9} (1988) 107.},\ref\hatf{B. Hatfield,
{\twelveit Quantum Field Theory
of Particles and Strings} (Addison Wesley, ??????, 1992).}}$, in which
states are described by wave functionals $\Psi(\phi)$
of a c-number field $\phi({\bf r})$,
the inner product is defined by functional integrals
\eqn\stinnpro{<\Psi_1|\Psi_2>
\sim \int D\phi  \Psi^*_1(\phi) \Psi_2(\phi) ~,}
the action of the field on the state is by multiplication
\eqn\stx{\Phi |\Psi> \sim \phi \Psi(\phi) ~,}
and the action of the canonical momentum on the state is by
functional differentiation
\eqn\stp{\Pi |\Psi>
\sim {1 \over i} {\delta \over \delta \phi} \Psi(\phi) ~.}
In this formalism one finds the nice result that if $\Psi_G(\phi)$
is a Gaussian functional of the form
\eqn\gau{\Psi_G(\phi)
\equiv \exp \biggl[ - {1 \over 4} \int d{\bf x}~d{\bf y} ~
(\phi({\bf x}) - \phi_0({\bf x})) ~ G^{-1}({\bf x},{\bf y}) ~
(\phi({\bf y}) - \phi_0({\bf y})) \biggr] ~,}
then the expectation of the Hamiltonian in $\Psi_G(\phi)$
automatically gives
\eqn\gaubub{<H>_G =
\int D\phi  \Psi_G^*(\phi) H \Psi_G(\phi) = E^{bubble}(\phi_0,G)~,}
where $E^{bubble}(\phi_0,G)$ is the approximation
of the effective energy for composites which includes only the $O(\lambda)$
contribution to $E_2(\phi_0,G)$.
Therefore by stationarizing $<H>_G$ with respect
to variations of $G$ one obtains the
``bubble-resummed" effective energy.

\vglue 0.3cm
\leftline{\twelveit 2.6. Comments and Cautions}
\vglue 1pt
Some caution is necessary in using the generating functionals that
we have discussed.
Let us start by analyzing some features of the
effective potential\footnote{*}{\ninerm\baselineskip=11pt
Clearly the issues discussed in this
section also concern the effective action
and the effective energy, but for definiteness
we base the presentation on the case of the effective potential.}
in the simple $\lambda \Phi^4$ scalar theory. Including only the classical
and the one loop term one has (see Sec.2.3)
\eqn\vfeu{\eqalign{V(\phi) \simeq V_{cl}(\phi)  +  {1 \over 2}
\int_k \ln [k^2 -a + {\lambda \over 2} \phi^2]  ~, }}
where we have set up a Euclidean evaluation of the one loop contribution.
(N.B. We are not doing Euclidean field theory. We simply evaluate
a Minkowski space-time integral using a
continuation to Euclidean space.)

The first important observation is that the integral in Eq.\vfeu\space is
clearly divergent, and therefore a regularization/renormalization
procedure is needed. It turns out
that this renormalization can be performed in a rather standard
fashion$^{\colewei-\wei}$.

Eq.\vfeu\space also
indicates that the effective potential is not necessarily
real and convex (if $a>0$ the effective potential in Eq.\vfeu\space
is not convex and has an imaginary
part for small $\phi$),
and therefore it cannot
be interpreted as the free energy of the system.
It has been suggested$^{\ref\colebook{COLEBOOK.. .. ... ... .... ...}}$ that
the imaginary part of the effective potential might
carry information on the decay properties of the system.
The possible non-convexity of the effective potential
is important in the Legendre transform
procedure$^{\ref\wiede{On the issue of convexity of the effective potential
also see the contribution of Wiedemann in these proceedings}}$;
in fact, it may not be possible to find a unique solution for
$J(\phi)$ in Eq.\fidef, because, {\elevenit e.g.}, there are multiple roots.
In these cases, if one is
interested in the free energy, a Maxwell construction
can be used, but then the resulting effective potential will not
be the generating functional for the Green's functions.

Concerning the effective potentials of gauge theories there is obviously an
additional issue to be addressed: gauge invariance.
In a gauge theory the effective potential, besides depending on a
background scalar field (and its complex conjugate), can also depend on a
background gauge field: $V(A_\mu,\phi,\phi^*)$.
There are two ways in which the effective potential
may fail to be gauge invariant.
First it may not be invariant against gauge transformations
of $A_\mu$, $\phi$, $\phi^*$.
Second, even if the effective potential is
invariant against gauge transformations
of $A_\mu$, $\phi$, $\phi^*$, it may still depend on the gauge
choice made in quantization; for example in covariant gauges
it may depend on the coefficient $\xi$ of the
gauge fixing term $\xi (\partial_\mu A^\mu)^2$.
In general the effective potential does depend on $\xi$.

The gauge dependence of the effective potential can be understood in terms
of the fact that off-shell Green's functions are not physical and are not
gauge invariant.
Although the full effective potential need not be a physical quantity
in a gauge theory,
gauge invariant physical information can be obtained from it, for example
at the stationary points.
However, in many cosmological models the detailed profile of
the effective potential, even away from the stationary points,
plays an important role (slow rolling down out of equilibrium).
Therefore for these models the physical
interpretation of the full effective potential is an important issue.

\vglue 0.6cm
\leftline{\twelvebf 3. Finite Temperature Relativistic
Quantum Field Theory}
\vglue 0.4cm
The observable properties of a relativistic quantum field theory
at nonzero temperature\footnote{*}{\ninerm\baselineskip=11pt
For a review, see Refs.[$\ref\vanwe{N. Landsman and C. van Weert,
Phys. Rep. {\twelvebf 145} (1987) 142.},
\ref\kapu{J. Kapusta, {\twelveit Finite Temperature
Field Theory} (Cambridge University Press,
Cambridge, 1989).}$].} can be extracted from the thermal Green's
functions, defined by
\eqn\tgf{\eqalign{G_\beta(x_1,.....,x_n) \equiv
< {\cal T} \Phi(x_1) ..... \Phi(x_n) >
= {tr ~ e^{- \beta H} T \Phi(x_1) ..... \Phi(x_n) \over
tr ~ e^{- \beta H}} ~ .}}

Let us consider the two-point thermal Green's function $G_\beta(x_1,x_2)$.
The Dyson-Schwinger equations of motion for $G_\beta(x_1,x_2)$
follow from the Heisenberg equations of motion like
in the ordinary zero temperature case
\eqn\tsd{\eqalign{\square_1 G_\beta(x_1,x_2) =
- i \delta(x_1-x_2) +
< {\cal T} \square \Phi(x_1) \Phi(x_2) >
{}~.}}
The structure of Eq.\tsd\space is exactly the same as in the zero temperature
case.
The difference between the
thermal Green's function and
the zero temperature Green's function
is merely in the boundary conditions,
which determine the unique, physically appropriate solution
to the differential equation.
One can therefore follow the same method of solution to Eq.\tsd\space
used at zero temperature, except for what concerns the boundary conditions.

For definiteness and simplicity,
we illustrate the methods of solution of Eq.\tsd\space
in the case of bosonic fields, for which the boundary conditions are
\eqn\boundcond{\eqalign{ G_\beta(x_1,x_2)|_{x_1^0=0}
= G_\beta(x_1 ,x_2)|_{x_1^0=- i \beta}
{}~.}}
These follow from the cyclicity of trace, which defines $G_\beta$,
see Eq.\tgf.
[Note that Eq.\tsd\space is in real time, but the solution
is required to be periodic in imaginary time.]
We also assume that, as it happens in most cases of physical interest,
translation invariance is not broken, and therefore
\eqn\gtrinv{\eqalign{ G_\beta(x_1,x_2) = G_\beta(x_1 - x_2,0)
\equiv G_\beta(x_1 - x_2)
{}~.}}
Because of the translation invariance
it is convenient to go to conjugate
space.

In the ``real time formalism"
one introduces the Fourier integral transform $G_\beta(p)$
of $G_\beta(x_1 - x_2)$:
\eqn\fouri{\eqalign{ G_\beta(x_1-x_2) =
\int_p e^{- i p (x_1-x_2)}
G_\beta(p)
{}~.}}
It then follows simply from the fact that the field $\Phi(x)$
is an operator which verifies causal commutation relations,
that $G_\beta(p)$
has the following integral representation
\eqn\intrap{\eqalign{ G_\beta(p) = & i
\int^{\infty}_{-\infty} {dp'_0 \over (2 \pi)}
{\rho_\beta(p'_0, {\bf p}) \over p_0 -p'_0 + i \epsilon}
+ f(p_0) \rho_\beta(p_0, {\bf p})
\cr
= & i
\int^{\infty}_{-\infty} {dp'_0 \over (2 \pi)}
\rho_\beta(p'_0, {\bf p}) \biggl[
{1+ f(p'_0) \over p_0 -p'_0 + i \epsilon} -
{f(p'_0) \over p_0 -p'_0 - i \epsilon} \biggr]
{}~,}}
where $f$ is the bosonic distribution function
\eqn\fbos{\eqalign{ f(E) = {1 \over e^{\beta E} - 1}
{}~,}}
and $\rho_\beta$ is a density function to be
determined using Eq.\tsd.

In the lowest order in perturbation theory the density function
is known,
\eqn\rhofree{\eqalign{\rho^{free}_\beta =
2 \pi \epsilon(p_0)
\delta(p^2-m^2) ~,}}
and correspondingly the thermal 2-point Green's function is given by
\eqn\tgfree{\eqalign{ G^{free}_\beta(p) = {i \over p^2-m^2 + i \epsilon} +
{2 \pi \over e^{\beta E} - 1} \delta(p^2-m^2)
{}~.}}
[Here $\epsilon(p_0) \equiv \theta(p_0)-\theta(-p_0)$, and $\theta$ is
the unit step function, vanishing for negative argument.]
Note that $G^{free}_\beta$ satisfies the free-field Schwinger-Dyson
equation $(p^2-m^2) G^{free} = i$, and the temperature dependence
is confined in the term, proportional to $\delta(p^2-m^2)$, in which
boundary conditions are encoded.

As an alternative to the real time formalism one can work in the ``imaginary
time formalism", in which the Green's functions are considered in
imaginary time.
As a consequence of the periodicity in imaginary time
described in Eq.\boundcond,
in this formalism the
momentum space two-point thermal Green's functions are the
$G_\beta(\omega_n,{\bf p})$ defined by
Fourier series and integral transforms
\eqn\matstras{\eqalign{ [G_\beta(x)]_{x_0=-i x_4}  =
{i \over \beta} \sum_n e^{- \omega_n x_4}
\int {d{\bf p}  \over (2 \pi)^3} e^{i {\bf p} {\bf x}}
G_\beta(\omega_n,{\bf p})
{}~,}}
where
\eqn\matsfreq{\eqalign{ \omega_n = i {2 \pi n \over \beta}
{}~.}}
In Eq.\matstras\space there is
a Fourier series for the time-like component
and an ordinary Fourier integral transform for the
space-like components.

The same analysis which leads to the spectral representation
of the real time Green's functions also leads to the following
spectral representation
for the imaginary time Green's functions:
\eqn\intrapim{\eqalign{ G_\beta(\omega_n,{\bf p}) = i
\int^{\infty}_{-\infty} {dp'_0 \over (2 \pi)}
{\rho_\beta(p'_0, {\bf p}) \over \omega_n -p'_0 }
{}~,}}
which involves
the same spectral function $\rho_\beta$ present in Eq.\intrap,
and the same denominator structure, but now $p_0$ is substituted by the
imaginary discrete quantity $\omega_n$ and therefore, because the denominator
never vanishes, there is
no need for $i \epsilon$-prescriptions.

The imaginary time Green's function which corresponds to
the spectral function $\rho^{free}_\beta$ is
\eqn\tgfreeim{\eqalign{ G^{free}_\beta(\omega_n,{\bf p})
= {i \over \omega_n^2-{\bf p}^2-m^2 }
{}~,}}
which looks like
a free zero temperature Green's function, except that
$p_0$ has been replaced
by $\omega_n$.

Perturbation theory has its most physical realization in the
real time formalism, however some technical problems are encountered.
The structure of perturbation theory is the same as at zero temperature
because the structure of the equations is the same,
but it is difficult to implement the boundary conditions.
Moreover, the Green's functions involve a $\delta$-function
and therefore in high order calculations one encounters ambiguous
products of
$\delta$-functions at the same point.
(Only the one-loop results can be obtained rather easily in the real time
formalism because no products of $\delta$-functions at the same point
can occur.)
Obviously these difficulties can be overcome by
doing things carefully, but it is very hard to
find algorithms for perturbative calculations in the
real time formalism.
A systematic algorithmic prescription
is given by the {\elevenit time path approach}$^{\vanwe}$, but
involves mathematical elaborations that obscure physical content:
a specific closed time contour in the complex plane is chosen, and
one uses both time- and anti-time-ordered
Green's functions with an effective doubling of the degrees of
freedom.

Perturbation theory is not pathologic
in the imaginary time formalism.
In fact,
one can introduce Feynman rules in complete analogy to
zero temperature, except for the
substitution in the propagators
of the continuous variable $p_0$
by the discrete $\omega_n$,
and the
substitution of the integration over a loop four-momentum
by an integration over the three spatial components
of the momentum and a sum over the discrete values of the
time-like component:
\eqn\imfeyn{\eqalign{\int_p
\rightarrow
{i \over \beta} \sum^{\infty}_{n=-\infty} \int {d{\bf p} \over (2 \pi)^3}
{}~.}}

Unfortunately the simplicity of perturbation theory
in the imaginary time formalism is somewhat offset by the fact that,
when one is interested in the answers to dynamical questions, it is
necessary to continue results back to real time.
However, the effective potential,
which is the generating functional for Green's functions at zero
momentum, can be computed in the imaginary time formalism
because at zero momentum (static physics)
there is no difference between real time and imaginary time.
For example for the $\lambda \Phi^4$ scalar theory both in
imaginary time and in real time one finds that the one-loop
effective potential is given by
\eqn\olepexample{V(\phi)=V_{cl}(\phi)+{1 \over \beta} \sum_n
\int {d{\bf k}  \over (2 \pi)^3}
\ln ({4 \pi^2 n^2 \over \beta^2} +{\bf k}^2 -a +
{\lambda \over 2} \phi^2)  ~.}

\vglue 0.6cm
\leftline{\twelvebf 4. Non-Equilibrium
Quantum Field Theory}
\vglue 0.3cm
\leftline{\twelveit 4.1. Isoentropic Energy-non-conserving Time Evolution}
\vglue 1pt
A physical system is described by its density matrix $\rho$,
\eqn\defro{\rho \equiv { e^{- \beta H} \over
tr ~ e^{- \beta H}} ~,}
and average values of observables $O$ are
determined by the density matrix:
\eqn\vevro{<O> = tr  \rho O  ~.}

Finite temperature field theory describes equilibrium physics,
and therefore involves a time-independent density matrix.
In general, however, the density matrix is time-dependent,
and the task of non-equilibrium quantum field theory is to study the
time evolution of $\rho$.

Non-equilibrium physics is a vast subject,
and there is no canonical approach to its investigation.
Usually the approach is suggested by the specifics
of the physical system that one wants to describe.
We discuss an approach which is
useful in early universe cosmology, and is set up in the
framework of the field theoretic Schr\"{o}dinger picture,
which is particularly suitable to time-dependent problems
that require an initial condition for specific solution.

The functional density matrix is given by a superposition
of wave functionals
\eqn\schroro{\rho(\phi_1,\phi_2) = \sum_n p_n \Psi_n(\phi_1)
\Psi^*_n(\phi_2) ~,}
were $\{ \Psi_n \}$ is a complete set of wave functionals,
and $p_n$ is the probability ($\sum_n p_n =1$)
that the system is in the state $\Psi_n$.
In general both $\Psi_n$ and $p_n$ are time-dependent.

In equilibrium the dynamics is time-translation invariant and
energy is conserved.
The complete set of wave functionals $\{ \Psi_n \}$ can be chosen
to be the set of
the (time-dependent) energy eigenstates,
and the $p_n$ are time-independent and are given by the canonical
Boltzmann distribution:
\eqn\pn{p_n = { e^{- \beta E_n} \over
\sum_n e^{- \beta E_n}} ~,}
where $E_n$ is the energy eigenvalue of the state $\Psi_n$.
The time evolution of the density matrix is trivial: it remains constant
in time because both the $p_n$'s  and $\Psi_n \Psi_n^*$
are constant (N.B. the time dependence of the $\Psi_n$'s is just a phase).

For non-equilibrium physics the time evolution of the density matrix is
instead nontrivial. In fact, the $p_n$'s need not be Boltzmann factors
and can change in time, and it might not be possible to choose
the $\Psi_n$'s as energy eigenstates.
We assume that the time dependence of the $\Psi_n$'s
be determined by the time-dependent Schr\"{o}dinger equation.
As a consequence, the density matrix $\rho$ satisfies the following
differential equation
\eqn\tero{{d \rho \over dt} = \sum_n p_n {d \over dt} (\Psi_n \Psi_n^*)
+\sum_n {d p_n \over dt} (\Psi_n \Psi_n^*)
= i [ \rho , H]
+\sum_n {d p_n \over dt} (\Psi_n \Psi_n^*)
{}~.}
In order for Eq.\tero\space to describe a well defined initial value
problem for the time evolution of $\rho$,
it is necessary to give the form of $H$ and a model
for $dp_n/dt$.
We investigate Eq.\tero\space in the case of time independent $p_n$'s
(which corresponds
to entropy-conserving time
evolution) and time dependent
Hamiltonian\footnote{*}{\ninerm\baselineskip=11pt
The physical idea behind these assumptions is that one drops
the time dependence of the $p_n$'s, in order to make the equations
treatable, but tries to account for (part of)
this time dependence
by introducing an ``effective" time-dependent Hamiltonian.}
(which corresponds to energy non-conserving
time evolution)$^{\ref\piebo{O. Eboli, R. Jackiw, and S.-Y. Pi,
Phys. Rev. {\twelvebf D37} (1988) 3557;
R. Jackiw, Physica {\bf A158} (1989) 269.}}$.
With these hypotheses Eq.\tero\space takes the form of the quantum
Liouville-von Neumann equation with time dependent Hamiltonian
\eqn\lvn{{d \rho \over dt} = i [ \rho , H]
{}~.}

In particular,
we assume that the ``interesting"
time dependence of $H$ occurs in an interval $t_i<t<t_f$ whereas
$H=H_i=constant$ when
$t<t_i$ and
$H=H_f=constant$ when $t>t_f$.
Interesting Hamiltonians, which may be investigated, include
Hamiltonians with time-dependent mass squared (in particular
the mass squared changing sign is useful in the study of
symmetry-changing phase transitions), and Hamiltonians
corresponding to a quantum field theory in a background
Robertson-Walker metric with time-dependent scale factor.

We impose as initial condition that the density
matrix $\rho$ when $t \le t_i$ be given by the Boltzmann distribution
for $H_i$ with temperature $T_i$.
The typical result, which one seeks using the formalism that we developed,
is the density matrix at times $t>t_f$,
{\elevenit i.e.} one solves Eq.\lvn\space and examines $\rho$ at late times.
If $\rho$ still changes with time when $t>t_f$, one concludes that the
system remains out of equilibrium.
If $\rho$ is time-independent for $t>t_f$,
there are two possibilities: either $\rho$ is given by a Boltzmann
distribution for $H_f$ with a temperature $T_f$, in which case
the system regains thermal equilibrium at temperature $T_f$,
or $\rho$ is not given by a Boltzmann distribution, in which case
the system reaches non-thermal equilibrium.

\vglue 0.3cm
\leftline{\twelveit 4.2. Methods for Solving the
Liouville-von Neumann equation}
\vglue 1pt
We are interested in obtaining
solutions to the Liouville-von Neumann equation \lvn\space
with a canonical density matrix as initial condition.
However, the exact solution of this problem is only
feasible$^{\piebo}$
for problems that are described by a quadratic Hamiltonian.
For more general Hamiltonians one can obtain approximate solutions
using the observation that the Liouville-von Neumann equation can be derived
from a variational principle introduced by Balian and
Veneroni$^{\ref\bave{BALIAN BEVERONI .........}}$.
An approximate application of this variational principle with a restricted
variational {\elevenit Ansatz} leads to approximate equations
for the density matrix that may be integrated.

The Liouville-von Neumann equation can be derived by varying the actionlike
quantity
\eqn\actlike{I=- \int^{t_f}_{t_i} dt~
tr \biggl[\rho \left( {d \Lambda \over dt} +
i [H,\Lambda] \right) \biggr] - [tr (\rho \Lambda)]_{t=t_i}
{}~,}
where $\Lambda$, $\rho$, and $H$ are time-dependent kernels and the trace
is over these kernels. $\Lambda$ is a Lagrangian multiplier kernel.
By varying $I$ with respect to $\Lambda$ and $\rho$ one obtains respectively
\eqn\varl{\delta_\Lambda I= \int^{t_f}_{t_i} dt~
tr \biggl[{d \rho \over dt} +
i [H,\rho] \biggr] \delta \Lambda - [tr (\rho \delta \Lambda)]_{t=t_f}
{}~,}
\eqn\varr{\delta_\rho I= - \int^{t_f}_{t_i} dt~
tr \biggl[{d \Lambda \over dt} +
i [H,\Lambda] \biggr] \delta \rho- [tr (\Lambda \delta \rho)]_{t=t_i}
{}~.}
As in all time dependent variational principles we impose boundary
conditions; we require that
\eqn\bla{[\delta \Lambda]_{t=t_f}= 0 ~,}
so we may set
\eqn\bl{[\Lambda]_{t=t_f}= identity
{}~.}
Also we require, according to our program,
\eqn\br{[\rho]_{t=t_i}= Boltzmann~distribution
{}~,}
and therefore
\eqn\br{[\delta \rho]_{t=t_i}= 0 ~.}
Demanding that $I$ be stationary against both variations
of $\Lambda$ and $\rho$, with
the above boundary conditions,
gives the Liouville-von Neumann
equation for $\rho$, and also for $\Lambda$.
The boundary condition Eq.\bl\space selects the static
solution $\Lambda=identity$, for all time,
so that $\Lambda$ disappears from the discussion
and we are left with a variational formulation of the Liouville-von Neumann
equation for $\rho$.

If, rather than performing arbitrary $\Lambda$ and $\rho$ variations,
we evaluate $I$ with specific parameter dependent expressions
for $\Lambda$ and $\rho$ and then vary these parameters, we obtain an
approximate solution of the problem.
In particular, a Gaussian {\elevenit Ansatz} can be very useful.
In this case $\rho$ can be parametrized, also exploiting the
Hermiticity of $\rho$ [$\rho(\phi_1,\phi_2)=\rho^*(\phi_2,\phi_1)$],
as
\eqn\gauro{\rho(\phi_1,\phi_2) =
e^{-\gamma}
exp \biggl\{ - {1 \over 2} \biggl[
\phi_1 \biggl( {G^{-1} \over 2} -2 i \Pi \biggr) \phi_1
+ \phi_2 \biggl( {G^{-1} \over 2} +2 i \Pi \biggr) \phi_2
- \phi_1 \biggl( G^{-1/2} \xi G^{-1/2} \biggr) \phi_2
\biggr]
\biggr\}
{}~,}
where $G$ and $\Pi$ are real and symmetric kernels, $\xi$ is Hermitian,
and $\gamma$ is to be chosen so that $\rho$ is properly normalized.
$\xi$ is called the {\elevenit degree of mixing}; in fact, it is a measure
of the amount by which $\rho$ differs from a pure state. For $\xi=0$,
$\rho=\Psi(\phi_1) \Psi^*(\phi_2)$, with
\eqn\purestate{\Psi(\phi) = (det^{-1/4} 2 \pi G)
exp \biggl[ - {1 \over 2} \phi \biggl( {G^{-1} \over 2} -2 i \Pi \biggr)
\phi \biggr]
{}~.}
The meaning of $G$ and $\Pi$ can be understood by looking at the following
bilinear expectation values
\eqn\meaningpg{\eqalign{<\Phi({\bf r}) \Phi({\bf r}')> = &
[G^{1/2} (1-\xi)^{-1} G^{1/2}]({\bf r},{\bf r}')
\cr
<\Pi({\bf r}) \Pi({\bf r}')> = &
{1 \over 4} [G^{-1/2} (1+\xi) G^{-1/2}]({\bf r},{\bf r}')
+ 4 [\Pi G^{1/2} (1-\xi)^{-1} G^{1/2}
\Pi]({\bf r},{\bf r}')
\cr
<\Phi({\bf r})  \Pi({\bf r}')> = &
{i \over 2} \delta({\bf r}-{\bf r}')+
2 [G^{1/2} (1-\xi)^{-1} G^{1/2}
\Pi]({\bf r},{\bf r}')
{}~.}}

This approximation scheme has been used in an investigation
of early universe evolution$^{\piebo}$.
The results support the validity of the heuristic
method
of analysis of early universe phase
transitions
(a dynamical problem) in which one evaluates
only the (static) effective potential and assumes
that the evolution be adequately described by substituting
the effective potential in place of the classical potential
in classical evolution equations$^{\ref\guthpi{GuthPi ... ... ..... ...}}$.

\vglue 0.6cm
\leftline{\twelvebf 5. Chern-Simons Theory}
\vglue 0.3cm
\leftline{\twelveit 5.1. Physical Relevance of 3-dimensional Quantum
Field Theories}
\vglue 1pt
In this section we discuss a particular 3-dimensional quantum field theory.
There are several physical circumstances
in which a 3-dimensional quantum field theory can be relevant.

A first example is given by Minkowski 4-dimensional field theory
in the
Schr\"{o}din- ger Hamiltonian formalism, wherein one works
at fixed time and therefore
the c-number arguments of the wave functionals
depend only on the three spatial coordinates,
they are functionals of 3-dimensional fields.

Another example is given by situations in which the interesting
physics is
(at least approximately) confined
to a plane.
This scenario occurs in some condensed matter phenomena,
like the quantum Hall effect and high $T_c$ superconductivity,
and for motion
in presence of
cosmic strings, which is described
by planar gravity.

Yet another circumstance in which
3-dimensional field theories are relevant is
directly related to the topic of this workshop.
In fact, one expects that
in the high temperature limit
thermal field theory in four dimensions can be
described
by a 3-dimensional field theory.
A naive formal argument, which is often used to support this
expectation, is based on the observation that,
because of the periodic boundary conditions imposed on the fields,
finite temperature field theory in the imaginary time formalism
``lives" in $R^3 \times S^1$, and the radius of the $S^1$
is given by $1/T$.
In the limit $T \rightarrow \infty$ the $S^1$ collapses leaving a
3-dimensional space $R^3$.
This apparent dimensional reduction in the high temperature limit
can be also seen in Feynman diagrams; for example
in the $\lambda \Phi^4$ scalar theory the tadpole
contribution to the self-energy is
\eqn\tdpse{i T \sum_n \int {d{\bf p} \over (2 \pi)^3}
{}~ \lambda ~
{i \over -4 n^2 \pi^2 T^2 - {\bf p}^2 - m^2}
{}~,}
and its high temperature limit, in which only the $n$=0 term of the sum
survives, is given by
\eqn\tdpseht{\int {d{\bf p}
\over (2 \pi)^3}  ~ \lambda T ~ {1 \over {\bf p}^2 + m^2}
{}~.}
The passage from
Eq.\tdpse\space to Eq.\tdpseht\space can be interpreted again as a
dimensional reduction (and a
dimensionful scaling of the coupling $\lambda \rightarrow \lambda T$).

This possible thermal field theory application
was one of the initial motivations for studying
the specific 3-dimensional field theory,
involving the Chern-Simons term, which we review in the following.
Recent results$^{\ref\nair{NAIR IN BANFF ***** ***** ***** *****}}$
in the context of the {\elevenit hard loops} high temperature approximation
have found that
a quantity which naturally appears in the context of Chern-Simons theories
is indeed directly related to the generating functional for hard thermal
loops in QCD.

\vglue 0.3cm
\leftline{\twelveit 5.2. The Chern-Simons Term }
\vglue 1pt
The Chern-Simons term\footnote{*}{\ninerm\baselineskip=11pt
For a review, see Ref.[$\ref\jakplan{R. Jackiw, in {\twelveit
Physics, Geometry, and Topology}, ed. H. C. Lee
(Plenum, New York, 1990).}$].}
$\Omega(A)$ is a structure which
can be constructed
out of gauge fields in three dimensions:
\eqn\csterm{\Omega(A) \equiv - {1 \over 8 \pi^2} \epsilon^{ijk}
tr (\partial_i A_j A_k + {2 \over 3} A_i A_j A_k)
{}~.}
(If the gauge fields are Abelian
the last term on the right is absent.)
$\Omega(A)$ is a topological quantity, it is independent
of the metric.

Even though it is a 3-dimensional object, the Chern-Simons term
appears in the study of some 4-dimensional problems.
It arose first in the Schr\"{o}dinger-picture description of
the vacuum angle of quantized 4-dimensional
non-Abelian gauge theories.
This comes about in the following manner. Consider the quantity
\eqn\wa{W(A) \equiv \int d{\bf x}~ \Omega(A)
{}~,}
which satisfies Gauss law, {\elevenit i.e.}
\eqn\wagau{\biggl(D_i {\delta \over \delta A_i} \biggr)_a W(A) =
D_i \biggl({1 \over 16 \pi^2} \epsilon^{ijk} F^a_{jk} \biggr) = 0
{}~.}
However $W(A)$ is not gauge invariant:
under a finite
gauge transformation, corresponding to the gauge group element $g$,
$W(A)$ changes
by the winding number $n_g$ of the gauge transformation:
\eqn\windeq{W(A^g)-W(A)=\int d{\bf x}~ [\Omega(A^g)-\Omega(A)]=
\int d{\bf x}~{1 \over 8 \pi^2} \epsilon^{ijk}
[\partial_i (\partial_j g ~g^{-1} A_k)+\omega(g)] = n_g
{}~,}
where
\eqn\omeg{\omega(g) \equiv {1 \over 24 \pi^2} \epsilon^{ijk}
tr (g^{-1} \partial_i g ~g^{-1} \partial_j g ~g^{-1} \partial_k g)
{}~,}
$n_g\equiv \int d{\bf x}~ \omega(g)$,
and we assume that the gauge field vanishes sufficiently
rapidly as $|{\bf x}| \rightarrow \infty$ so that
the surface term, which arises from
$\int d{\bf x} ~ \epsilon^{ijk} \partial_i (\partial_j g g^{-1} A_k)$,
also vanishes.
It is also true that $\omega(g)$ can
be written as a total derivative, {\elevenit i.e.}
there exist some $\omega^i(g(q))$,
where $q(x)$ is a parametrization for $g(x)$, such that
\eqn\omegtotder{\omega(g) = \partial_i \omega^i(g(q))
{}~.}
However, because of the specific behavior of
the parametrization $q$ of
the group elements
as $|{\bf x}| \rightarrow \infty$, the integral of
$\omega(g)$ over the whole volume is not zero,
in spite of the fact that, owing to Eq.\omegtotder,
it can be presented as an integral over the surface at infinity.
Rather the integral takes an integer
value $n_g$,
which characterizes the homotopic equivalence class to which $g$ belongs.

The relevance of all this to 4-dimensional QCD can be seen in the
Schr\"{o}dinger
representation, where states are 3-dimensional
functionals that must satisfy Gauss law.
Therefore they can take the form
\eqn\qcdstat{\Psi(A)=e^{i \theta W(A)} \psi(A)
{}~,}
where $\psi(A)$ is gauge invariant,
and consequently obeys Gauss law.
The states $\Psi(A)$ also
satisfy Gauss law
but they
are not gauge invariant.
Rather under a gauge transformation $g$, belonging to the homotopy
class $n_g$, they change according to
\eqn\jakadd{\Psi(A) \rightarrow \Psi(A^g) = e^{- i \theta n_g} \Psi(A)
{}~,}
This is the origin of the famous vacuum $\theta$-angle.

In 3-dimensional physics, the Chern-Simons term can be added to the
ordinary Lagrange density  of Yang-Mills theories
leading to the Lagrange density of the
{\elevenit topologically massive theories}, which is given by
\eqn\ymcs{{\cal L}_m = {1 \over 2} tr F^{\mu \nu} F_{\mu \nu}
+ 8 \pi^2 k \Omega(A)
{}~.}
${\cal L}_{m}$ is not gauge invariant; however, at the classical
level ${\cal L}_{m}$ describes a gauge invariant theory
because the equations of motion that follow
from ${\cal L}_{m}$ are gauge invariant:
\eqn\ymeqmo{D_\mu F^{\mu \nu} + {k \over 2} \epsilon^{\nu \alpha \beta}
F_{\alpha \beta} = 0
{}~.}
At the quantum
level, the quantity that defines the field theory is the phase exponential
of the action, and one sees$^{\jakplan}$ that it is gauge invariant
for all values of $k$ when the gauge group is Abelian, whereas in the
non-Abelian case gauge invariance requires that $k$ be quantized
as follows: $4 \pi k = integer$.

Interestingly, one can show that
the gauge bosons described by ${\cal L}_{m}$ have mass $k$
(therefore these theories, like the Higgs models, are counterexamples
to the naive statement
that gauge invariance requires gauge bosons to be massless),
and this is
the reason for the name topologically
massive theories.

$\Omega(A)$ is also related to the Pontryagin density $P$.
In fact,
\eqn\pontry{P= - {1 \over 32 \pi^2} tr \epsilon^{\mu \nu \alpha \beta}
F_{\mu \nu} F_{\alpha \beta} = \partial_\alpha \Omega^\alpha(A)
{}~,}
where
\eqn\omealpha{\Omega^\alpha(A) = - {1 \over 8 \pi^2}
\epsilon^{\alpha\mu \nu \beta}
tr (\partial_\mu A_\nu A_\beta + {2 \over 3} A_\mu A_\nu A_\beta)
{}~,}
and clearly, for fixed $\alpha$, $\Omega^\alpha(A)$ can
be identified with $(\pm) \Omega(A)$ if the
dependence of the gauge fields on the $\alpha$-th coordinate
is suppressed.

Numerous \space mathematically \space interesting \space
applications \space have \space been \space found \space for \space
the

\noindent
Chern-Simons term;
for example, the correlators of Wilson lines
in a pure Chern-Simons theory (the theory discussed in the
following subsection)
are related to the polynomial invariants of knot
theory$^{\ref\witt{WITTEN}}$.

\vglue 0.3cm
\leftline{\twelveit 5.3. Pure Chern-Simons Theory}
\vglue 1pt
The mathematical structure
which is important
in the context of the hard loops high temperature approximation
arises in {\elevenit pure Chern-Simons theory}.
The Lagrange density ${\cal L}_{CS}$
of this theory is obtained as the $k \rightarrow \infty$
limit of ${\cal L}_{m}$, {\elevenit i.e.} by neglecting in ${\cal L}_m$
the Yang-Mills kinetic term.
Therefore ${\cal L}_{CS}$ is given by
\eqn\pcs{{\cal L}_{CS} = 8 \pi^2 k \Omega(A)
{}~.}
The equation of motion which follows from ${\cal L}_{CS}$ is
\eqn\emclpcs{\epsilon^{\alpha \mu \nu} F_{\mu \nu} = 0 ~.}

Classically this theory is trivial: the solutions of the equation
of motion satisfy $F_{\mu \nu} = 0$ and they are therefore pure gauge.
However, as a quantum field theory this simple model can have some
interesting structure.
We analyze the theory as a canonical quantum field theory in
the gauge $A_0 =0$; the Lagrange density is
\eqn\pcsao{{\cal L}_{CS} = {k \over 2} \epsilon^{ij} tr {\dot A}_i A_j
{}~.}
By varying
with respect to $A_i$ one obtains the equations of motion
\eqn\pcsaoem{{\dot A_i} =0
{}~.}
The Hamiltonian vanishes: $H=0$.
Because of the choice $A_0 =0$,
the equation gotten from Eq.\pcs\space
by varying
with respect to $A_0$ is not obtained;
rather a constraint is imposed
\eqn\constr{G^a \equiv - {k \over 2} \epsilon^{ij} F^a_{i j} =0
{}~.}
This is the present analog of the Gauss law constraint in ordinary
Yang-Mills theory. (One easily verifies that the $G^a$'s
are the generators
of static gauge transformations.)

There are two ways of quantizing this theory: one can either solve the
constraint \constr\space first and then quantize the remaining degrees
of freedom or quantize first and then solve
the constraint.
If one solves the constraint first
the vector potentials are pure gauges,
and if the topology of the space
is trivial
no interesting structure arises in the theory.
We follow the alternative strategy, {\elevenit i.e.} quantize
before solving the constraint, which does lead
to some interesting structures.

The Lagrange density ${\cal L}_{CS}$ involves only
first order time derivatives;
for this reason, unlike what happens in Yang-Mills theories,
the components of the vector potential do not commute:
\eqn\commut{[A_{i}^a({\bf x}),A_{j}^b({\bf y})] = {i \over k}
\epsilon_{ij} \delta({\bf x}-{\bf y})
{}~.}

Since $H=0$, there is no dynamics, and all the interesting aspects
of the theory come from the solution of the constraint,
{\elevenit i.e.} by requiring that the generators $G^a({\bf x})$
annihilate the physical states
\eqn\gaugau{G^a({\bf x}) |\Psi> =0 ~.}

We solve Eq.\gaugau\space in the Schr\"{o}dinger picture.
Because the spatial components of the vector potential do not commute,
the wave functionals depend on just one of the two spatial components.
[Instead for 3-dimensional Yang-Mills theories in the (2-dimensional)
fixed-time Schr\"{o}dinger picture,
the wave functionals depend on both spatial components.]
We choose $A^a_1$, which we call $\phi^a$,
as the argument of the the wave functionals $\Psi(\phi)$,
and $A^a_2$ is realized by functional differentiations with
respect to $\phi^a$,
\eqn\carte{\eqalign{|\Psi> \sim & \Psi(\phi) \cr
A_{1}^a({\bf x}) |\Psi> \sim & \phi^a({\bf x}) \Psi(\phi) \cr
A_{2}^a ({\bf x}) |\Psi> \sim & {1 \over i k}
{\delta \over \delta \phi^a({\bf x})} \Psi(\phi)
{}~,}}
which is
consistent with the commutation relations in Eq.\commut.

In our realization of the Schr\"{o}dinger picture,
the constraint
equation \gaugau\space can be written in the form of functional
differential equation
\eqn\gaugauschro{\biggl( \partial_1
{\delta \over \delta \phi^a({\bf x})} +
f^{a b c} \phi^b({\bf x})
{\delta \over \delta \phi_c({\bf x})}
- i k \partial_2 \phi^a({\bf x}) \biggr) \Psi(\phi) = 0
{}~.}

We solve the Gauss law constraint in two steps. First we determine
the result of a finite
gauge transformation $g$, implemented by the unitary operator
$U(g)$, on a state, {\elevenit i.e.} we evaluate
\eqn\determ{U(g) \Psi(\phi) = e^{i \int_x \lambda^a(x) G^a(x)} \Psi(\phi)
\equiv e^{i G} \Psi(\phi)
{}~,}
[in Eq.\determ\space we also indicate the explicit form of
$U(g)$ in terms of the generators $G^a$, and
thereby define the operator $G$],
and then we demand that
\eqn\demand{U(g) \Psi(\phi) =  e^{i G} \Psi(\phi)
=\Psi(\phi)
{}~,}
as required by the Gauss law constraint $G^a |\Psi> =0$.

Let us observe that
\eqn\ggfi{\eqalign{G \equiv & \int_x \lambda^a G^a
= - \int_x \lambda^a \biggl( \partial_1 {1 \over i}
{\delta \over \delta \phi^a} + f^{abc} \phi^b
{\delta \over \delta \phi_c} \biggr)
- k \int_x \phi^a \partial_2 \lambda^a
\equiv G_\phi + 2 k \int_{\bf x} tr \phi \partial_2 \lambda
{}~,}}
where $G_\phi$ is the generator of infinitesimal gauge transformations
on the argument $\phi^a=A_1^a$ of the wave functional.
Due to the presence of
the term $2 k \int_{\bf x} tr \phi \partial_2 \lambda $,
the full action of a gauge transformation on the state is realized with
a 1-cocycle (see Ref.[\ref\duntrujak{G. Dunne,
R. Jackiw, and C. Trugenberger,
Phys. Rev. {\bf D9} (1974) 1686.}]):
\eqn\cocy{U(g) \Psi(\phi) \equiv e^{i G} \Psi(\phi) =
e^{i G} e^{- i G_\phi} e^{i G_\phi} \Psi(\phi)
=e^{i G} e^{- i G_\phi} \Psi(\phi^g)
= e^{- 2 \pi i \alpha_1(\phi;g)} \Psi(\phi^g)
{}~,}
where $\phi^g \equiv g^{-1} \phi g + g^{-1} \partial_1 g$, and
$\alpha_1$ is obtained by evaluating $e^{i G} e^{- i G_\phi}$
with the help of the Baker-Hausdorff
procedure
\eqn\cocyalpha{\alpha_1(\phi;g) \equiv
- {k \over 2 \pi} \int_x
tr (2 \phi \partial_2 g ~g^{-1}
+ g^{-1} \partial_1 g ~g^{-1} \partial_2 g)
+ 4 \pi k \int_x \omega^0(g)
{}~.}
The $\omega^0$ which appears in Eq.\cocyalpha\space is the time-like
component of the 3-vector $\omega^\mu$, which we already encountered
in Eq.\omegtotder, and one can obtain an explicit expression for $\omega^0$
only after a parametrization for $g(x)$ has been chosen.
Alternatively $\int_x \omega^0(g)$ may be presented as the integral
of $\omega(g) \equiv \partial_\mu \omega^\mu(g)$
over a three-manifold $M$, whose boundary is the two-space on which our
Schr\"{o}dinger functional is defined.

\noindent
$\alpha_1(\phi;g)$ is a one-cocycle because it satisfies the condition
\eqn\onecocy{\alpha_1(\phi;g) = \alpha_1(\phi;g {\tilde g})
-\alpha_1(\phi^g;{\tilde g})
{}~,}
which is called the one-cocycle condition.

Eq.\cocy\space concludes the first step in our strategy for
solving the Gauss law constraint; at this point
we demand that $U(g) \Psi(\phi) = \Psi(\phi)$, and find that
the physical wave functionals are not gauge invariant, but rather
they satisfy
\eqn\blablo{\Psi(\phi^g) = e^{2 \pi i \alpha_1(\phi;g)} \Psi(\phi)
{}~.}
A general wave functional which verifies Eq.\blablo\space can be written
in the form
\eqn\solpagsix{\Psi(\phi) = e^{i W(\phi)} \psi(\phi)
{}~.}
where
$\psi(\phi)$ is completely
gauge invariant [$\psi(\phi^g)=\psi(\phi)$],
$W(\phi)$ is given by
\eqn\solalpha{W(\phi) \equiv
- k \int_x
tr ( h^{-1} \partial_1 h \, h^{-1} \partial_2 h)
+ 8 \pi^2 k \int_x \omega^0(h)
{}~,}
and $h$ is to be expressed in terms of $\phi$ using
$\phi \equiv h^{-1} \partial_1 h$.
$W(\phi)$ is a {\elevenit cochain}, it satisfies
$W(\phi) =  W(\phi^g)- 2 \pi \alpha_1(\phi;g)$.

The term $4 \pi k \int_x \omega^0(h)$ is multivalued, and as a consequence
it turns out that the in order for the wave functional
in Eq.\solpagsix\space to be single valued one needs to require
that $4 \pi k$=integer.
This is the way that the quantization condition,
which we obtained by requiring the gauge invariance of the phase exponential
of the action, arises in the present, Schr\"{o}dinger-picture
formalism.

Note that on a topologically trivial space, one cannot
construct a gauge invariant functional depending only on $\phi^a=A_1^a$.
Consequently $\psi(\phi)$ in Eq.\solpagsix\space is a constant $N$, and the
general state of Chern-Simons theory is just
\eqn\jakaddtwo{\Psi(\phi) = N e^{i W(\phi)}
{}~.}

In conclusion, it is interesting to verify that Eq.\constr\space has indeed
been satisfied. To this end we consider $A_i \Psi(\phi)$.
For $i=1$, we have the parametrization (by definition)
$A_1=\phi=h^{-1} \partial_1 h$.
Hence $A_1 \Psi(\phi)=h^{-1} \partial_1 h \Psi(\phi)$.
For $i=2$, $A_2=(1/i) \delta/\delta\phi^a$ and
$A_2^a \Psi(\phi)= (\delta W(\phi)/\delta\phi^a) \Psi(\phi).$
While $W$ is presented in Eq.\solalpha\space
as a functional of $h$, we may use the chain rule, and the above formula
relating $\phi$ to $h$.
In this way one finds $A_2(\phi) \Psi(\phi)=h^{-1} \partial_2 h \Psi(\phi)$
and one concludes that, acting on states, $A_i$ is a pure gauge
\eqn\jakaddtre{A_i \Psi(\phi)=h^{-1} \partial_i h \Psi(\phi)
{}~.}
Therefore, the corresponding
field strength vanishes, in keeping with Eq.\constr.

\vglue 0.3cm
\leftline{\twelveit 5.4. The Chern-Simons Eikonal }
\vglue 1pt
In 1+1-dimensional point-particle quantum mechanics with Lagrangian
\eqn\actpp{L = {m {\dot x}^2 \over 2} - V(x)
{}~,}
and energy constraint
\eqn\eneco{{p^2 \over 2 m} + V(x) = E
{}~,}
the {\elevenit eikonal}, which is the exponent of the WKB wave functions,
is given by
\eqn\eikpm{W(x) \equiv \int^x dx'~p(x')
{}~,}
where $p(x)$ is the solution of Eq.\eneco: $p(x) =  \sqrt{2m[E - V(x)]}$.

In Chern-Simons theory the Lagrangian can be written as
\eqn\leik{L_{CS} = k~ tr {\dot A}_1 A_2
{}~,}
the Hamiltonian vanishes, and there is the constraint
\eqn\coneik{F^a_{12}=
\partial_1 A_2^a - \partial_2 A_1^a + f^{abc} A_1^b A_2^c =0
{}~.}
Clearly there is an analogy with the
1+1-dimensional point-particle quantum mechanics based on the
identifications $A_1 \sim x$ and $k A_2 \sim p$.
One can therefore introduce WKB wave functional
\eqn\wfeik{\Psi(\phi) =
\exp \left(i \int^{\phi^a} DA_1~ k ~A_2^a(A_1) \right)
\equiv \exp i W(\phi)
{}~,}
where $A_2^a(A_1)$ is to be obtained by solving Eq.\coneik.

An explicit solution $A_2^a(A_1)$ of Eq.\coneik\space is obtained
perturbatively in Ref.[\nair] as a series in $A_1$.
Here, we observe that $W(\phi)$
coincides with the phase exponential of the wave functional that
solves the Gauss law constraint (Eqs.\solpagsix-\solalpha),
{\elevenit i.e.}
the WKB/eikonal wave functional is exact in Chern-Simons theory,
owing to the simple dynamical structure of that theory.
To see the result, observe that $W(\phi)$ defined by Eq.\wfeik\space
satisfies
\eqn\weab{{\delta W(\phi) \over \delta \phi^a} = k \, A_2^a ~.}
and as a consequence of Eq.\coneik, $W(\phi)$ also satisfies
\eqn\gaueiko{\partial_1 {\delta W(\phi) \over \delta \phi^a({\bf x})} +
f^{a b c} \phi^b({\bf x})
{\delta W(\phi) \over \delta \phi_c({\bf x})}
- k \partial_2 \phi^a({\bf x}) = 0 ~.}
This shows that the wave functional in Eq.\wfeik\space
satisfies Eq.\gaugauschro.

\vglue 0.6cm
\leftline{\twelvebf Acknowledgements}
\vglue 0.4cm
This work is
supported in part by funds provided by the U.S. Department of Energy (D.O.E.)
under contract \#DE-AC02-76ER03069.

\vglue 0.6cm
\leftline{\twelvebf References}
\vglue 0.4cm
\itemitem{\jona.} G. Jona-Lasinio, {\twelveit Nuovo Cimento}
{\twelvebf 34} (1964) 1790.
\itemitem{\colewei.} S. Coleman and E. Weinberg,
{\twelveit Phys. Rev.} {\twelvebf D7} (1973) 1888.
\itemitem{\jakdi.} R. Jackiw, {\twelveit Phys. Rev.} {\twelvebf D9}
(1974) 1686.
\itemitem{\doja.} L. Dolan and R. Jackiw,
{\twelveit Phys. Rev.} {\twelvebf D9} (1974) 3320.
\itemitem{\wei.} S. Weinberg, {\twelveit Phys. Rev.}
{\twelvebf D9} (1974) 3357.
\itemitem{\rive.} R. J. Rivers, {\twelveit Path Integrals Methods in Quantum
Field Theory} (Cambridge University Press, Cambridge, 1987).
\itemitem{\corn.} J. M. Cornwall, R. Jackiw, and E. Tomboulis,
{\twelveit Phys. Rev.} {\twelvebf D10} (1974) 2428.
\itemitem{\gacpi.} G. Amelino-Camelia and S.-Y. Pi,
{\twelveit Phys. Rev.} {\twelvebf D47} (1993) 2356.
\itemitem{\timedep.} A. Kerman and
R. Jackiw, {\twelveit Phys. Lett.} {\twelvebf A71}
(1979) 159;
R. Jackiw, {\twelveit Int. J. Quantum Chem.} {\twelvebf 17} (1980) 41.
\itemitem{\schrpict.} R. Jackiw, {\twelveit Canad. Math.
Soc. Conference Proceedings}
{\twelvebf 9} (1988) 107;
{\twelveit Field Theory in Particle
Physics}, ed. O.J.P. Eboli, M. Gomez,
and A. Santoro (World Scientific, Singapore, 1990).
\itemitem{\hatf.} B. Hatfield, {\twelveit Quantum Field Theory
of Particles and Strings} (Addison Wesley, Redwood City, 1992).
\itemitem{\colebook.} S. Coleman, {\twelveit Aspects of
Symmetry} (Cambridge University Press, Cambridge, 1985).
\itemitem{\wiede.} On the issue of convexity of the effective potential
also see the contribution of U. Wiedemann in these proceedings.
\itemitem{\vanwe.} N. Landsman and C. van Weert,
{\twelveit Phys. Rep.} {\twelvebf 145} (1987) 142.
\itemitem{\kapu.} J. Kapusta, {\twelveit Finite Temperature
Field Theory} (Cambridge University Press, Cambridge, 1989).
\itemitem{\piebo.} O. Eboli, R. Jackiw, and S.-Y. Pi,
{\twelveit Phys. Rev.} {\twelvebf D37} (1988) 3557;
R. Jackiw, {\twelveit Physica} {\twelvebf A158} (1989) 269.
\itemitem{\bave.} R. Balian and M. Veneroni,
{\twelveit Phys. Rev. Lett.} {\twelvebf 47} (1981) 1353;
{\twelvebf 47} (1981) 1765(E);
{\twelveit Ann. Phys.} {\twelvebf 164} (1985) 334.
\itemitem{\guthpi.} For a review, see L. F. Abbott and S.-Y. Pi,
{\twelveit Inflationary Cosmology} (World Scientific, Singapore, 1986).
\itemitem{\nair.} See the contribution of V.P. Nair in these proceedings.
\itemitem{\jakplan.} R. Jackiw, in {\twelveit
Physics, Geometry, and Topology}, ed. H. C. Lee
(Plenum Press, New York, 1990).
\itemitem{\witt.} E. Witten,
{\twelveit Comm. Math. Phys.} {\twelvebf 121} (1989) 351.
\itemitem{\duntrujak.} G. Dunne, R. Jackiw, and C. Trugenberger,
{\twelveit  Ann. Phys.} {\twelvebf 149} (1989) 197.

\end